\documentclass{article}
\usepackage{amsmath,amsthm,amssymb}
\usepackage[a4paper]{geometry}
\usepackage{graphicx}
\usepackage[textwidth=8em,textsize=normalsize]{todonotes}
\usepackage{natbib}
\usepackage{lipsum}
\usepackage{float}
\usepackage{subcaption}
\usepackage{dcolumn}
\usepackage{url}
\usepackage{xcolor}
\usepackage{multirow}
\usepackage[linesnumbered,ruled]{algorithm2e}

\newcommand{\mtx}[1]{\mathbf{#1}}
\newcommand{\vct}[1]{\mathbf{#1}}

\def \mA {\mtx{A}}
\def \mC {\mtx{C}}

\def \mI {\mtx{I}}

\def \mM {\mtx{M}}

\def \mQ {\mtx{Q}}

\def \mS {\mtx{S}}
\def \mU {\mtx{U}}

\def \mX {\mtx{X}}

\def \mbeta {\boldsymbol{\beta}}

\def \mdelta {\boldsymbol{\delta}}
\def \mepsilon {\boldsymbol{\varepsilon}}
\def \meta {\boldsymbol{\eta}}
\def \mmu {\boldsymbol{\mu}}

\def \mgamma {\boldsymbol{\gamma}}

\def \mSigma {\boldsymbol{\Sigma}}

\def \mLambda {\mtx{\Lambda}}
\def \mOmega {\mtx{\Omega}}

\def \mPhi {\mtx{\Phi}}

\def \zero     {\mathbf{0}}
\def \mzero    {\vct{0}}
\def \mone    {\vct{1}}
\def \ma {\vct{a}}

\def \mm {\vct{m}}

\def \ms {\vct{s}}

\def \mx {\vct{x}}
\def \my {\vct{y}}

\def \N {\mathcal{N}}

\def \I {\mathcal{I}}
\def \G {\mathcal{G}}
\def \B {\mathcal{B}}
\def \T {\mathcal{T}}
\def \U {\mathcal{U}}

\usepackage{template_tex}

\newcommand\blfootnote[1]{%
  \begingroup
  \renewcommand\thefootnote{}\footnote{#1}%
  \addtocounter{footnote}{-1}%
  \endgroup
}

\begin{document}

\title{A Fully Bayesian Approach for Comprehensive Mapping of Magnitude and Phase Brain Activation in Complex-Valued fMRI Data}
\author{Zhengxin Wang \\
Clemson University
\and
Daniel B. Rowe \\
Marquette University
\and
Xinyi Li \\
Clemson University
\and
D. Andrew Brown \blfootnote{\emph{Address for correspondence}: 
D. Andrew Brown, School of Mathematical and Statistical Sciences, Clemson University, Clemson, SC, USA. Email: ab7@clemson.edu} \\
Clemson University}

\date{} 

\maketitle

\begin{abstract}
Functional magnetic resonance imaging (fMRI) plays a crucial role in neuroimaging, enabling the exploration of brain activity through complex-valued signals. These signals, composed of magnitude and phase, offer a rich source of information for understanding brain functions. Traditional fMRI analyses have largely focused on magnitude information, often overlooking the potential insights offered by phase data. In this paper, we propose a novel fully Bayesian model designed for analyzing single-subject complex-valued fMRI (cv-fMRI) data. Our model, which we refer to as the CV-M\&P model, is distinctive in its comprehensive utilization of both magnitude and phase information in fMRI signals, allowing for independent prediction of different types of activation maps. We incorporate Gaussian Markov random fields (GMRFs) to capture spatial correlations within the data, and employ image partitioning and parallel computation to enhance computational efficiency. Our model is rigorously tested through simulation studies, and then applied to a real dataset from a unilateral finger-tapping experiment. The results demonstrate the model's effectiveness in accurately identifying brain regions activated in response to specific tasks, distinguishing between magnitude and phase activation.
\end{abstract}

\vspace{9pt}


\section{Introduction}
Magnetic resonance imaging (MRI) is a non-invasive imaging technique that has revolutionized the field of medical diagnostics and research, particularly in the realm of neuroimaging. Functional magnetic resonance imaging (fMRI) is a subtype of MRI specifically optimized for higher temporal resolution. While conventional MRI captures static anatomical details, fMRI extends the utility by enabling the examination of metabolic functions over time. It has become indispensable in a variety of applications ranging from diagnosis of pathological conditions to the investigation of complex physiological processes in the human brain.

\begin{figure}
\begin{center}
\includegraphics[width=5in]{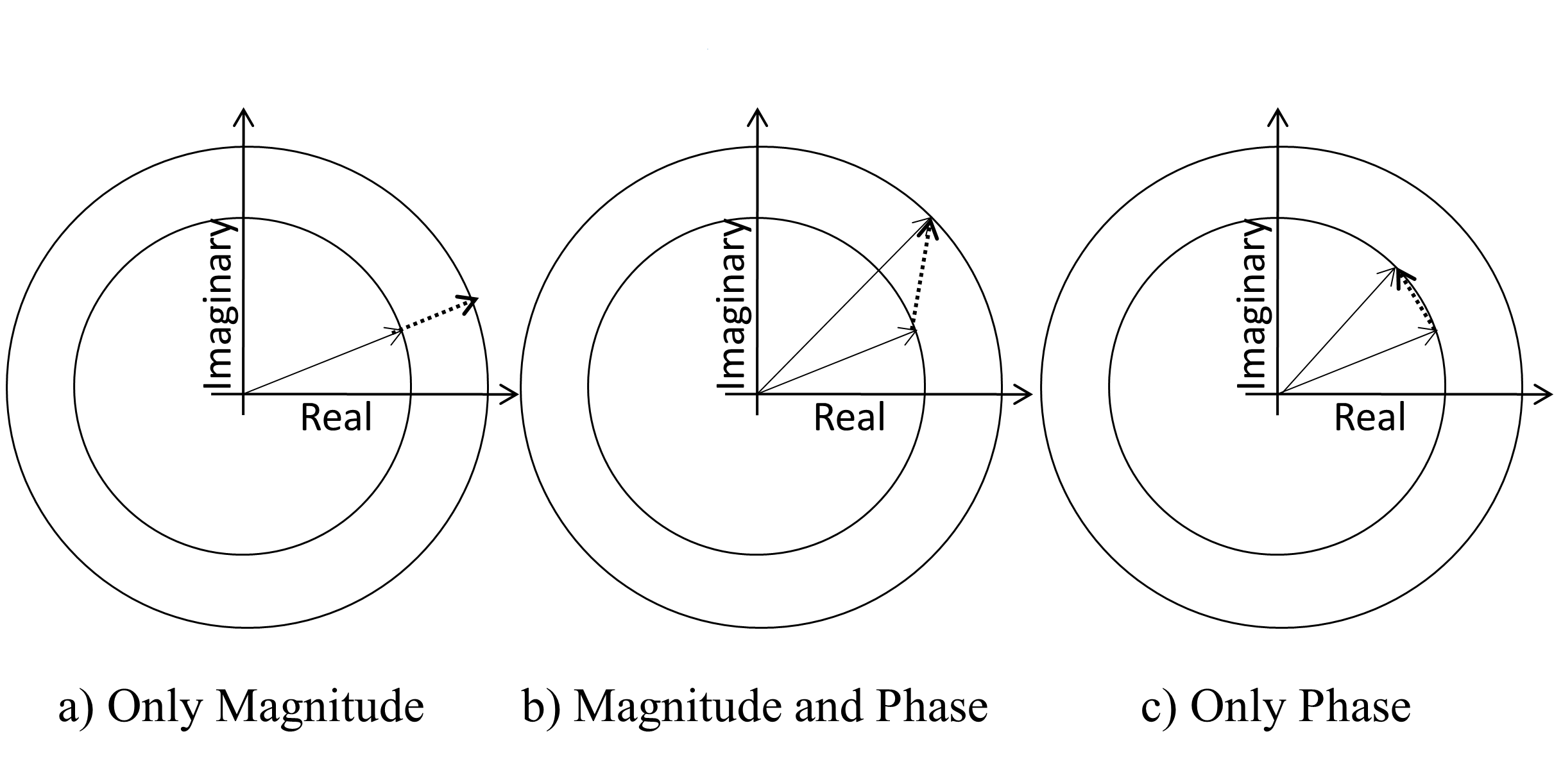}
\end{center}
\caption{Illustration of potential changes in complex-valued fMRI time series \citep{Rowe2019}.}
 \label{fig:changes}
\end{figure}

FMRI inherently generates complex-valued signals characterized by real and imaginary components, and further summarized as magnitude and phase. This complex structure arises from the forward and inverse Fourier transformations executed in the data collection process, which are affected by phase imperfections \citep{Brown2014}. These signals may exhibit changes in magnitude, phase, or both over time in response to a stimulus, as shown in Figure~\ref{fig:changes}.

The magnitude changes in complex-valued fMRI (cv-fMRI) are fundamentally driven by the blood-oxygenation-level-dependent (BOLD) effect, which operates through a cascade of hemodynamic responses. Neuronal activity leads to increased demand of oxygen, so the freshly oxygenated blood fluxes into the active region and displaces deoxygenated blood, leading to an overall increase in the oxygenation level of the blood in that region. These changes in blood oxygenation cause a change in BOLD signal and magnetic susceptibility, affecting the magnitude of MR signal. Thus, the BOLD effect can be considered an indirect measure of neuronal activity, mediated through vascular changes \citep{Boynton1996, Logothetis2008}.

On the other hand, phase changes are influenced not only by the BOLD effect but also by the electrical neuronal activities directly. These activities generate moving charges, and therefore create magnetic field which changes the phase of MRI signal. For this reason, the phase changes are able to reveal the aspects of neuronal activity or other phenomena that might be undetected by magnitude-based analyses \citep{Petridou2006}. By accurately modeling these phase changes, researchers can gain insights into the more direct effects of neuronal activity on the MRI signal, potentially leading to more precise and informative interpretations of fMRI data \citep{Feng09}. This is especially crucial in understanding complex brain functions and improving the accuracy of fMRI in research and clinical applications.

Traditionally, fMRI studies that aim to map brain activity have predominantly focused on analyzing only the magnitude of these MR signals \citep{Friston1994, Lindquist2008}. The phase components are frequently disregarded during the preprocessing steps. The magnitude-only analytical framework has its limitations. The first major limitation is that the omission of phase data results in the underutilization of valuable information that could be pertinent to understanding neurophysiological mechanisms. The second limitation, particularly relevant in studies that employ linear modeling \citep{Friston1994, Lindquist2008}, concerns the statistical assumptions made during the identification of active voxels (volumetric pixels in the imaging data). In such analyses, the expected BOLD response is usually modeled by the convolution of a ``boxcar'' binary stimulus function with either a gamma or double-gamma hemodynamic response function (HRF), then a voxel is identified as ``active'' if the magnitude of its complex-valued fMRI signal significantly varies with the expected BOLD response. This practice assumes that the error terms in the models are normally distributed. However, while the original real and imaginary components may follow a normal distribution, the magnitude actually adheres to a Ricean distribution that approximates a normal distribution only when the signal-to-noise ratios (SNRs) are sufficiently large \citep{Rice1944, Gudbjartsson1995}. Given that large SNRs are not universally guaranteed in fMRI studies, this statistical assumption becomes less reliable and exacerbates the limitation, consequently diminishing the power and reliability of the analysis.

In contrast, emerging research utilizing cv-fMRI data offers a more nuanced and comprehensive approach. By incorporating both the magnitude and phase of the MR signals \citep{Rowe2004, Rowe2005a, Rowe2005c, Rowe2005b, Rowe2007, Rowe2009a, Adrian2018}, or modeling real and imaginary components that both contain magnitude and phase information \citep{Lee2007, Yu2018}, cv-fMRI studies pave the way for the development of more robust and statistically powerful models. These models are better to handle variations in SNR and can fully exploit the available data, thereby offering potentially deeper and more accurate insights into task-related neuronal activity.

To accurately determine task-related brain activation maps from fMRI signals, fully Bayesian approaches have garnered attention due to their capacity to effectively model both spatial and temporal correlations. However, existing implementations of fully Bayesian methods in fMRI analysis have demonstrated notable shortcomings. For instance, certain studies have applied the fully Bayesian approach only to magnitude data \citep{Woolrich2004, Musgrove2016}, leading to underutilization of available data and flawed statistical assumptions as previously discussed. Additionally, others have employed fully Bayesian methods on cv-fMRI data, yet relied on a Cartesian model \citep{Yu2023, Wang2023}, which is limited to identifying active voxels without providing specific insights into the type of activation, be it in terms of magnitude, phase, or a combination of both.

In this paper, we propose a novel fully Bayesian model for mapping brain activity using single-subject cv-fMRI time series. Our model is designed to determine which voxels exhibit significant fMRI signal changes in response to a particular task, specifying the type and strength of these changes. This proposed Bayesian approach for fMRI data analysis is distinctive in its comprehensive utilization of both the real and imaginary components of fMRI data. It is capable of independently predicting different types of activation maps, in terms of magnitude, or phase, or both, capturing spatial correlations, and ensuring computational efficiency.

To achieve this, our approach incorporates Gaussian Markov random fields \cite[GMRFs;][]{RueHeld05} to effectively capture spatial associations present in cv-fMRI data. Moreover, we enhance computational efficiency by employing image partitioning and parallel computation strategies in our Markov chain Monte Carlo \citep[MCMC;][]{GelfandSmith90} algorithms.

The paper is structured as follows: Section \ref{Section: Model} introduces our proposed model, its parameters, and brain parcellation strategy. Section \ref{section: MySim} presents simulation studies and compares our approach with existing methods. In Section \ref{RealStudy}, we apply our model to a real finger-tapping experiment dataset. Section \ref{Conclusion} summarizes our findings and suggests future research directions.


\section{Model}\label{Section: Model}
In this section, we present our model designed for mapping brain activity using cv-fMRI data. Additionally, we introduce the brain parcellation strategy, which facilitates the parallel computation. Following this, we detail the implementation of a GMRF prior that effectively captures the spatial correlations inherent within the fMRI data. Finally, we describe an MCMC algorithm for approximating the posterior distribution of the parameters of interest.


\subsection{Model Formulation}
The polar model of \cite{Rowe2005c} has gained significant attention for modeling complex-valued fMRI data. Originating from the initial formulation with dynamic magnitude and constant phase \citep{Rowe2004}, the model has undergone several iterations \citep{Rowe2005a, Rowe2005b} to arrive at its current version to model both dynamic magnitude and dynamic phase. For a certain voxel $v$ (where $v=1, \hdots, V$) at time $t$ (where $t=1, \hdots, T$), its real and imaginary parts of complex-valued fMRI signal, $y_{v, Rt}$ and $y_{v, It}$, can be modeled as:
\begin{equation}\label{Rowesmodel}
    \begin{pmatrix}
        y_{v, Rt}\\
        y_{v, It}
    \end{pmatrix}
    =
    \begin{pmatrix}
        \rho_{v, t}\cos{\theta_{v, t}}\\
        \rho_{v, t}\sin{\theta_{v, t}}
    \end{pmatrix}
    +
    \begin{pmatrix}
        \varepsilon_{v, Rt}\\
        \varepsilon_{v, It}
    \end{pmatrix}
    , \quad \begin{pmatrix}
        \varepsilon_{v, Rt}\\
        \varepsilon_{v, It}
    \end{pmatrix} \sim \N(\mzero, \sigma_v^2\mI_2),
\end{equation}
where $\rho_{v, t}$ and $\theta_{v, t}$ are temporally varying magnitude and phase given by:
\begin{equation}
    \begin{split}
        \rho_{v, t}&=\beta_{v, 0}+x_t\beta_{v, 1},\\
        \theta_{v, t}&=\gamma_{v, 0}+u_t\gamma_{v, 1},\\
    \end{split}
\end{equation}
where $x_t$ and $u_t$ are the expected BOLD response and neuronal electromagnetic signal, respectively, at time $t$. Thus, for all time points:
\begin{equation}\label{Wangsmodel}
    \my_v=\begin{pmatrix}
        \mA_{R}(\mgamma_v)\\
        \mA_{I}(\mgamma_v)
    \end{pmatrix}
    \mX\mbeta_v+
    \mepsilon_v, \quad \mepsilon_v\sim \N(\zero, \sigma_v^2\mI_{2T}),
\end{equation}
where $\my_v=\left[(\my_{v,R})', (\my_{v,I})'\right]'\in \mathbb{R}^{2T}$ stacks real and imaginary components of cv-fMRI signal, and $\mX=[\mone, \mx]\in \mathbb{R}^{T \times 2}$ is the design matrix for the magnitude composed of ones and expected BOLD response $\mx$. The matrices $\mA_{R}(\mgamma_v), \mA_{I}(\mgamma_v)\in \mathbb{R}^{T \times T}$ are diagonal as:
\begin{equation}
\begin{split}
    \mA_{R}(\mgamma_v) &= 
\begin{pmatrix}
  \cos(\gamma_{v, 0} + u_1\gamma_{v, 1}) & 0 & \cdots & 0 \\
  0 & \cos(\gamma_{v, 0} + u_2\gamma_{v, 1}) & \cdots & 0 \\
  \vdots & \vdots & \ddots & \vdots \\
  0 & 0 & \cdots & \cos(\gamma_{v, 0} + u_t\gamma_{v, 1})
\end{pmatrix},\\
\mA_{I}(\mgamma_v) &= 
\begin{pmatrix}
  \sin(\gamma_{v, 0} + u_1\gamma_{v, 1}) & 0 & \cdots & 0 \\
  0 & \sin(\gamma_{v, 0} + u_2\gamma_{v, 1}) & \cdots & 0 \\
  \vdots & \vdots & \ddots & \vdots \\
  0 & 0 & \cdots & \sin(\gamma_{v, 0} + u_t\gamma_{v, 1})
\end{pmatrix},
\end{split}
\end{equation}
with a more compact form:
\begin{equation}
     \mA_{R}(\mgamma_v)=\text{diag}\left[\cos{\left(\mU\mgamma_v\right)}\right], \quad \mA_{I}(\mgamma_v)=\text{diag}\left[\sin{\left(\mU\mgamma_v\right)}\right],\\
\end{equation}
where $\mU=[\mone, \mtx{u}]\in \mathbb{R}^{T \times 2}$ is the design matrix for the phase composed of ones and neuronal electromagnetic signal $\mtx{u}$. Therefore, $\mbeta_v=\left[\beta_{v, 0}, \beta_{v, 1}\right]'\in \mathbb{R}^2$ and $\mgamma_v=\left[\gamma_{v, 0}, \gamma_{v, 1}\right]'\in \mathbb{R}^2$ are magnitude- and phase-related regression coefficients, respectively. The voxel-specific error term $\mepsilon_v$ follows a multivariate normal distribution with the variance-covariance matrix $\sigma_v^2\mI_{2T}$, and a Jeffreys prior can be assigned to $\sigma_v^2$ as $p(\sigma_v^2) \propto 1/\sigma_v^2$.


\subsection{Brain Parcellation and Spatial Priors}
Spatial correlations are a notable characteristic in fMRI signal data. Given that voxels represent an artificial segmentation of the brain's structure, they frequently display behaviors that are closely aligned with adjacent voxels \citep{Rowe2009b, Nencka2009, Karaman2014, Rowe2019}. To effectively model these spatial dependencies, it is beneficial to incorporate spatial structuring in the priors of $\mgamma_v$ and $\mbeta_v$ or in the hyperparameters of these priors. Moreover, a strategy of brain parcellation is applied to facilitate the parallel computation.


\subsubsection{Brain parcellation} In the study by \cite{Musgrove2016}, a technique for brain parcellation was introduced, focusing on the identification of active voxels within individual parcels before integrating these findings into a comprehensive map of brain activity. This approach involves dividing brain images into parcels, each containing around 500 voxels. By processing each parcel independently using an identical model and method, the technique allowed for parallel computation, enhancing computational efficiency. Similarly, \cite{Wang2023} adopted a comparable approach but differed in their strategy of dividing the brain into parcels of roughly equal geometric size. Both studies demonstrated that this parcellation strategy effectively minimizes edge effects, ensuring that the classification of border voxels in each parcel remains largely unaffected. Following the methodology of \cite{Wang2023}, we partitioned two- or three-dimensional fMRI images into a set number, $G$, of parcels, each of approximately equal geometric size. The choice of $G$ is based on empirical judgment, and as indicated by both \cite{Musgrove2016} and \cite{Wang2023}, variations within a reasonable range of $G$ do not significantly impact the results.


\subsubsection{Prior distributions of $\mbeta_v$ and $\mgamma_v$} For each parcel $g$ (where $g=1,\hdots,G$) encompassing $V_g$ voxels, we classify a voxel $v$ (where $v=1,\hdots,V_g$) based on its activity. Specifically, a voxel is classified magnitude-active if $\beta_{v, 1}\neq 0$, and phase-active if $\gamma_{v, 1}\neq 0$. Adhering to the spike-and-slab prior \citep{Mitchell1988}, the model is expressed as follows:
\begin{equation}
    \begin{split}
        \mbeta_v\mid\lambda_v, \tau_g^2&\sim\lambda_v\N_2\left(\mzero, \tau_g^2\mI\right)+(1-\lambda_v)\N_2\left(\mzero, \begin{pmatrix}
            \tau_g^2 & 0\\
            0 & 0
        \end{pmatrix}\right),\\
        \mgamma_v\mid\omega_v, \xi_b^2&\sim\omega_v\N_2\left(\mzero, \xi_g^2\mI\right)+(1-\omega_v)\N_2\left(\mzero, \begin{pmatrix}
            \xi_g^2 & 0\\
            0 & 0
        \end{pmatrix}\right).
    \end{split}
\end{equation}
In this formulation, $\lambda_v, \omega_v\in\{0, 1\}$ indicate the status of voxel $v$: $\lambda_v=1$ for a magnitude-active voxel and $\omega_v=1$ for a phase-active voxel, with 0 indicating inactivity in respective domains. The parameters $\tau_g^2$ and $\xi_g^2$ represent parcel-specific variances. These variances are constant for all voxels within a particular parcel but may vary across different parcels, and are assigned a Jeffreys prior, that is, $p(\tau_g^2) \propto 1/\tau_g^2$ and $p(\xi_g^2) \propto 1/\xi_g^2$, for $g = 1, \ldots, G$. The prior distributions can be succinctly represented as:
\begin{equation}
    \begin{split}
        \mbeta_v\mid\lambda_v, \tau_g^2&\sim\N_2\left(\mzero, \tau_g^2\mLambda_v\right),\quad \text{where }\mLambda_v=\begin{pmatrix}
        1&0\\
        0&\lambda_v
    \end{pmatrix},\\
    \mgamma_v\mid\omega_v, \xi_g^2&\sim\N_2\left(\mzero, \xi_g^2\mOmega_v\right),\quad \text{where }\mOmega_v=\begin{pmatrix}
        1&0\\
        0&\omega_v
    \end{pmatrix}.
    \end{split}
\end{equation}


\subsubsection{Spatial prior on $\lambda_v$ and $\omega_v$} To capture both spatial dependencies and the sparsity of active voxels in brain imaging, we implement a prior distribution for $\lambda_v$ and $\omega_v$. This approach is rooted in the hypothesis that voxels are more likely to mirror the activity (active/inactive) of their neighboring voxels \citep{Friston1994, Smith2007}, and there should be only a few active voxels across the entire brain from a simple task experiment \citep{Rao1996, Epstein1998}. We employ the sparse spatial generalized linear mixed model (sSGLMM) prior, formulated by \cite{Reich2006}, and later adopted by \cite{Hughes2013, Musgrove2016, Wang2023}. For voxel $v$ (where $v=1, ..., V_g$) within parcel $g$ (where $g=1, ..., G$), from the perspective of the magnitude, we suppose that:
\begin{equation}
    \begin{split}
        \lambda_v\mid\eta_{\lambda, v}&\stackrel {iid}\sim{\B}ern\left\{\mPhi(\psi_{\lambda}+\eta_{\lambda, v})\right\},\\
        \eta_{\lambda, v}\mid\mdelta_{\lambda, g}&\sim\N_1\left(\mm_v'\mdelta_{\lambda, g}, 1\right),\\
        \mdelta_{\lambda, g}\mid\kappa_{\lambda, g}&\sim\N_q\left\{\mzero, (\kappa_{\lambda, g}\mM_g{'}\mQ_g\mM_g)^{-1}\right\},\\
        \kappa_{\lambda, g}&\sim{\G}amma\left(a_{\kappa}, b_{\kappa}\right).
    \end{split}
\end{equation}
From the perspective of the phase:
\begin{equation}
    \begin{split}
        \omega_v\mid\eta_{\omega, v}&\stackrel {iid}\sim{\B}ern\left\{\mPhi(\psi_{\omega}+\eta_{\omega, v})\right\},\\
        \eta_{\omega, v}\mid\mdelta_{\omega, g}&\sim\N_1\left(\mm_v'\mdelta_{\omega, g}, 1\right),\\
        \mdelta_{\omega, g}\mid\kappa_{\omega, g}&\sim\N_q\left\{\mzero, (\kappa_{\omega, g}\mM_g{'}\mQ_g\mM_g)^{-1}\right\},\\
        \kappa_{\omega, g}&\sim{\G}amma\left(a_{\kappa}, b_{\kappa}\right).
    \end{split}
\end{equation}
In this sSGLMM prior, $\mPhi(\cdot)$ represents the cumulative distribution function (CDF) of the standard normal distribution. The terms $\psi_{\lambda}, \psi_{\omega} \in \mathbb{R}$ are fixed tuning parameters. The terms $\eta_{\lambda, v}, \eta_{\omega, v} \in \mathbb{R}$ are auxiliary parameters for the probit link functions. Spatial dependencies are modeled through constructs derived from the adjacency matrix $\mA_g$ of parcel $g$. This matrix,  $\mA_g \in \{0, 1\}^{V_g \times V_g}$, specifies neighborhood relations among voxels, with $\mA_{g,uv}=1$ indicating neighboring voxels $u$ and $v$ (based on user-defined criteria, typically voxels sharing an edge or a corner), and 0 otherwise. The matrix $\mM_g \in \mathbb{R}^{V_g \times q}$ is composed of the first $q$ principal eigenvectors of $\mA_g$. The row vector $\mm_v' \in \mathbb{R}^{1\times{q}}$ is the $v^{\text{th}}$ row of $\mM_g$, which is called ``synthetic spatial predictors'' \citep{Hughes2013}. The matrix $\mQ_g\in \mathbb{R}^{{V_g}\times {V_g}}$ is the graph Laplacian, that is, $\mQ_g=\text{diag}(\mA_g{\mone}_{V_g})-\mA_g$. The vectors $\mdelta_{\lambda, g}, \mdelta_{\omega, g}\in \mathbb{R}^{q\times1}$ are spatial random effects, and $\kappa_{\lambda, g}, \kappa_{\omega, g}\in \mathbb{R}$ are spatial smoothing parameters.

This sSGLMM prior captures the spatial correlations by using GMRFs, and introduces the sparsity by the selective inclusion of eigenvectors in $\mM_g$. Both \cite{Musgrove2016} and \cite{Wang2023} show it is good to capture the spatial correlations in fMRI data that align well with the parcellation strategy. In our simulation studies, the best-performing values of $\psi_{\lambda}$ and $\psi_{\omega}$ are pre-selected from a candidate list. For the real data, \cite{Musgrove2016} suggests the initial setting of $\mPhi^{-1}(0.02)=-2.05$, with later adjustments based on previous experiments' active voxel proportions. Moreover, we adopt $q=5$ (when $V_g\approx 200$), as shown by \cite{Hughes2013} that $q$ can be remarkably smaller than $V_g$. The Gamma distribution parameters, $a_{\kappa}=\frac{1}{2}$ and $b_{\kappa}=2000$ are the same for both magnitude and phase, leading to a large mean for $\kappa_{\lambda, g}$ and $\kappa_{\omega, g}$ ($a_{\kappa}b_{\kappa}$=1000), minimizing the risk of detecting spurious activity due to noise or other confounding factors.


\subsection{MCMC Algorithm and Posterior Distributions}\label{sbusec:MCMC}
We employ Gibbs sampling to obtain the joint and marginal conditional distributions of parameters of interest. Only the full conditional posterior distribution of $\mgamma_v$ is accessed via the Metropolis–Hastings algorithm \citep{Metropolis1953, Hastings1970}, the others follow known and available distributions. Detailed derivations and the required full conditional distributions are provided in the online supplementary material. To assess the convergence of the algorithm, we adopt the fixed-width diagnostic technique suggested by \cite{Flegal2008}. Convergence is considered achieved when the Monte Carlo Standard Error (MCSE) for all $\lambda_v$ and $\omega_v$ drops below 0.05, leading us to run $10^3$ iterations. After discarding the burn-in phase, the means of the sampled parameters are taken as point estimates. If $\widehat{\lambda}_v>0.925$, the voxel is magnitude-active; if $\widehat{\omega}_v>0.925$, it is phase-active. \cite{Smith2007} proposed the threshold of 0.8722 regarding the significance level $\alpha=0.05$. Since our approach is similar to a two-step sequential test, we use Bonferroni correction to make $\alpha=0.05/2=0.025$, leading to the adjustment of threshold from 0.8722 to 0.925.


\section{Simulation Studies}\label{section: MySim}
This section presents two distinct simulation studies. The first study focuses on a single map that comprises three types of active regions: one region is solely magnitude-active, another is solely phase-active, and the third is both magnitude- and phase-active. The second study involves multiple datasets, each containing only one type of activation on their maps. For comparative evaluation, we consider the following models:

\begin{itemize}
\item The model proposed by \cite{Musgrove2016}, referred to as MO, models magnitude-only data. For a certain voxel $v$, $v=1, \hdots, V$, over time $T$:
\begin{equation}\label{Musgrovemodel}
\begin{aligned}
\my_{v, M}=
\mX
\mbeta_{v, M}
+
\mepsilon_v,
\end{aligned}
\qquad
\begin{aligned}
\mepsilon_v\sim\N(\mzero, \sigma_v^2\mI_T)
\end{aligned}
\end{equation}
where $\my_{v, M}\in \mathbb{R}^{T}$ is the magnitude of complex-valued fMRI signal, and $\mX=[\mone, \mx]\in \mathbb{R}^{T \times 2}$ is the design matrix composed of ones and expected BOLD response $\mx$. The vector $\mbeta_{v, M}=(\beta_{v, M_0}, \beta_{v, M_1})'$ are regression coefficients.
\item The model delineated by \cite{Wang2023}, based on \cite{Lee2007}'s Cartesian model and referred to as CV-R\&I, models complex-valued data by modeling the real and imaginary components:
\begin{equation}\label{Leesmodel}
\begin{aligned}
\my_v=
\begin{pmatrix}
\mX & \mzero\\
\mzero & \mX
\end{pmatrix}
\begin{pmatrix}
\mbeta_{v, R}\\
\mbeta_{v, I}
\end{pmatrix}
+
\mepsilon_v, \quad \mepsilon_v\sim \N(\zero, \sigma_v^2\mI_{2T}),
\end{aligned}
\end{equation}
where $\my_v=\left[(\my_{v,R})', (\my_{v,I})'\right]'\in \mathbb{R}^{2T}$ is the stack of real and imaginary components of cv-fMRI signal. The vectors $\mbeta_{v, R}=(\beta_{v, R_0}, \beta_{v, R_1})'$ and $\mbeta_{v, I}=(\beta_{v, I_0}, \beta_{v, I_1})'$ are regression coefficients regarding real and imaginary components of cv-fMRI signal, respectively.
\item The model \eqref{Wangsmodel}, referred to as CV-M\&P, is based on \cite{Rowe2005c}'s polar model and models complex-valued data while characterizing magnitude and phase.
\end{itemize}

All three models adhere to a fully Bayesian approach, employ the sSGLMM spatial prior with brain parcellation strategy, and utilize Gibbs sampling to approximate their respective posterior distributions. The number of parcels $G$ is set to 16 for all models. Other tuning parameters, such as $\psi=\Phi^{-1}(0.35)$ for MO, $\psi=\Phi^{-1}(0.30)$ for CV-R\&I, and $\psi=\omega=\Phi^{-1}(0.42)$ for CV-M\&P, are predetermined to optimize prediction accuracy. The thresholds for identifying active voxels are set at 0.8722 for MO and CV-R\&I, as specified in their work, while CV-M\&P employs a threshold of 0.925, in accordance with Section~\ref{sbusec:MCMC}.

All results are generated by running the code on a custom-built desktop computer with an Intel Core i9-9980XE CPU (3.00GHz, 3001 Mhz, 18 cores, 36 logical processors), NVIDIA GeForce RTX 2080 Ti GPU, 64 GB RAM, and operating on Windows 10 Pro.


\subsection{Single Simulation}\label{subsection: Single simulation}


\subsubsection{Designed stimulus and expected BOLD response} 
\begin{figure}
    \begin{center}
        \includegraphics[width=\textwidth]{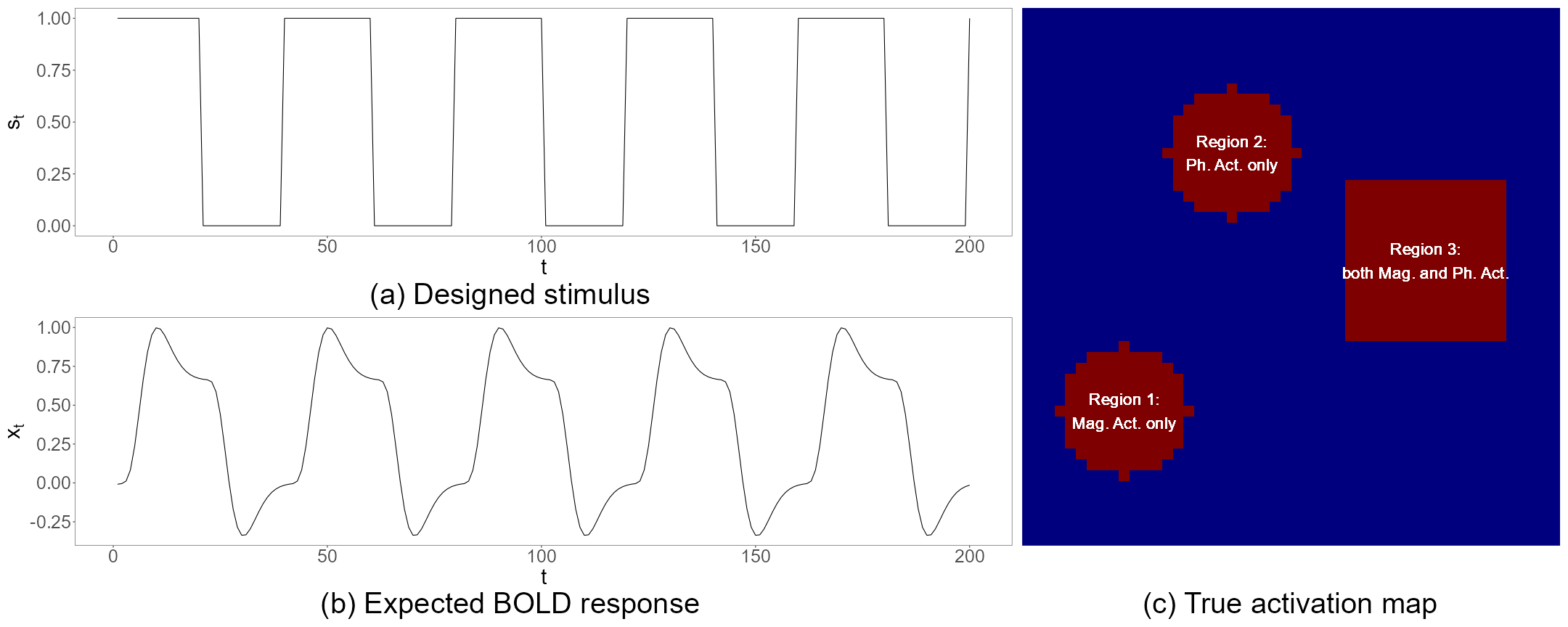}
    \end{center}
    \caption{(a) Designed stimulus; (b) Expected BOLD response; (c) True activation map.}
    \label{fig:True_maps}
\end{figure}

The designed stimulus $\ms$ is a binary signal comprised of five repeated epochs, each spanning 40 time points, resulting in a total duration of $T=200$ time points. Each epoch features the stimulus being alternately active and inactive, with both states persisting for 20 time points. We model the expected BOLD response $\mx$ by convolving this stimulus with a double-gamma HRF. Illustrations of both the designed stimulus and the expected BOLD response are provided in Figures~\ref{fig:True_maps}a and \ref{fig:True_maps}b, respectively, and are consistently used across all our simulation datasets.

\subsubsection{True activation map and true strength map} The true activation map contains three active regions on a $50\times 50$ panel, comprising two circles and one square, each with a radius of five. The exact locations of these regions are depicted in Figure~\ref{fig:True_maps}c. We want to assign distinct types of activation to each region: region 1 exhibits only magnitude activation, region 2 exhibits only phase activation, and region 3 exhibits both magnitude and phase activation, corresponding to the types illustrated in Figures~\ref{fig:changes}a, \ref{fig:changes}c, and \ref{fig:changes}b, respectively.

Utilizing the \texttt{specifyregion} function in the \texttt{neuRosim} library \citep{Welvaert2011} in \texttt{R} \citep{Rstats}, we initially generate a strength map with decay rates of 0.05, 0.05, and 0.15 for the three regions, respectively. This setup ensures that the central voxel of each active region has a strength of one, diminishing to zero towards the edges at the specified decay rate. For the true magnitude strengths, indicative of voxel response in magnitude to the stimulus, we multiply the strengths in regions 1 and 3 by 0.04909 and nullify the strengths in region 2, as represented in Figure~\ref{fig:MySim_estimated_maps}c. Similarly, for the true phase strengths, reflective of voxel response in phase, we multiply the strengths in regions 2 and 3 by a factor of $\pi/36$ and reduce the strengths in region 1 to zero, as illustrated in Figure~\ref{fig:MySim_estimated_maps}g. This methodology ensures that each region's activation profile is accurately mapped according to its designated stimulus response type.


\subsubsection{Simulating fMRI signals} We then simulate data according to Eq. \eqref{MySim_sim}:
\begin{equation}\label{MySim_sim}
    \begin{split}
         y_{v, Rt}&=(\beta_{0}+x_t\beta_{v, 1}){\rm cos}(\gamma_{0}+u_t\gamma_{v, 1})+\varepsilon_{v, Rt}, \quad \varepsilon_{v, Rt}\sim\N(0, \sigma^2),\\
        y_{v, It}&=(\beta_{0}+x_t\beta_{v, 1}){\rm sin}(\gamma_{0}+u_t\gamma_{v, 1})+\varepsilon_{v, It}, \quad \varepsilon_{v, It}\sim\N(0, \sigma^2),
    \end{split}
\end{equation}
where $\beta_{0} = 0.4909$, $\gamma_{0} = \pi/4$, and $\sigma = 0.04909$ are set constant for all voxels, and $x_t$ is the expected BOLD response $\mx$ from Figure~\ref{fig:True_maps}b at time $t$. It should be noted that we also use $\mtx{u}=\mx$ as the regressor for phase here when generating the data, but it could be its own neuronal electromagnetic signal $\mtx{u}$ for the phase in some cases. The signal-to-noise ratio for the magnitude (\text{SNR$_\text{Mag}$}) is thereby fixed at $\beta_{0}/\sigma=10$. The true values of $\beta_{1}$ and $\gamma_{1}$ generated previously in Figures~\ref{fig:MySim_estimated_maps}c and \ref{fig:MySim_estimated_maps}g are used, yielding the contrast-to-noise ratios for magnitude ($\text{CNR}_\text{Mag}$) and phase ($\text{CNR}_\text{Ph}$) as detailed in Eq. \eqref{MySim_CNR}:
\begin{equation}\label{MySim_CNR}
    \begin{split}
        \text{CNR}_\text{Mag}&=(\max{\beta_{v, 1}})/\sigma=0.04909/0.04909=1, \\
        \text{CNR}_\text{Ph}&=(\max{\gamma_{v, 1}})/{\text{SNR}_\text{Mag}}=(\pi/36)/10.
    \end{split}
\end{equation}


\subsubsection{Results} 
\begin{figure}
    \begin{center}
        \includegraphics[width=\textwidth]{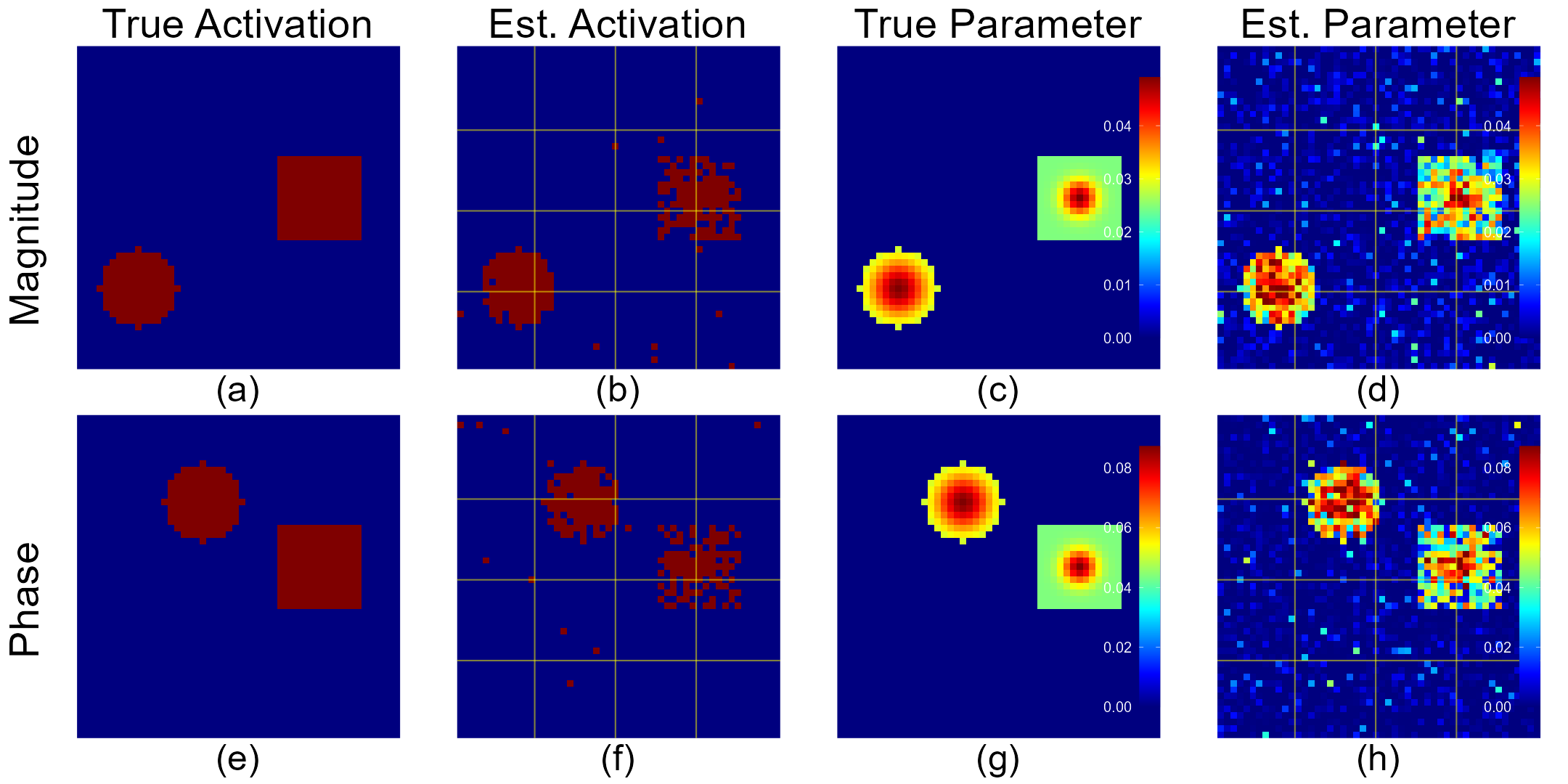}
    \end{center}
    \caption{(a) and (e) are true magnitude and phase activation maps; (b) and (f) are estimated activation maps as derived from CV-M\&P; (c) and (g) are true $\beta_1$ and $\gamma_1$; (d) and (h) are estimated $\beta_1$ and $\gamma_1$ as derived from CV-M\&P.}
    \label{fig:MySim_estimated_maps}
\end{figure}

Figure~\ref{fig:MySim_estimated_maps} presents both the true and estimated activation maps for magnitude and phase as derived from the CV-M\&P model, alongside the corresponding true and estimated parameters $\beta_1$ and $\gamma_1$. Notably, CV-M\&P effectively identifies separate regions that are active in magnitude and phase, and provides proper estimates for the parameters $\beta_1$ and $\gamma_1$. In the estimated activation maps (Figures~\ref{fig:MySim_estimated_maps}b and \ref{fig:MySim_estimated_maps}f), the overlap in the predicted active regions corresponds to the square-shaped region 3 in the true map (Figure~\ref{fig:True_maps}c), which is characterized by both magnitude and phase activation. When the predicted region 3 is excluded from these estimated maps, the remaining areas align well with the circular regions 1 and 2 in Figure~\ref{fig:True_maps}c, representing solely magnitude-active and solely phase-active voxels, respectively.

\begin{figure}
    \begin{center}
        \includegraphics[width=\textwidth]{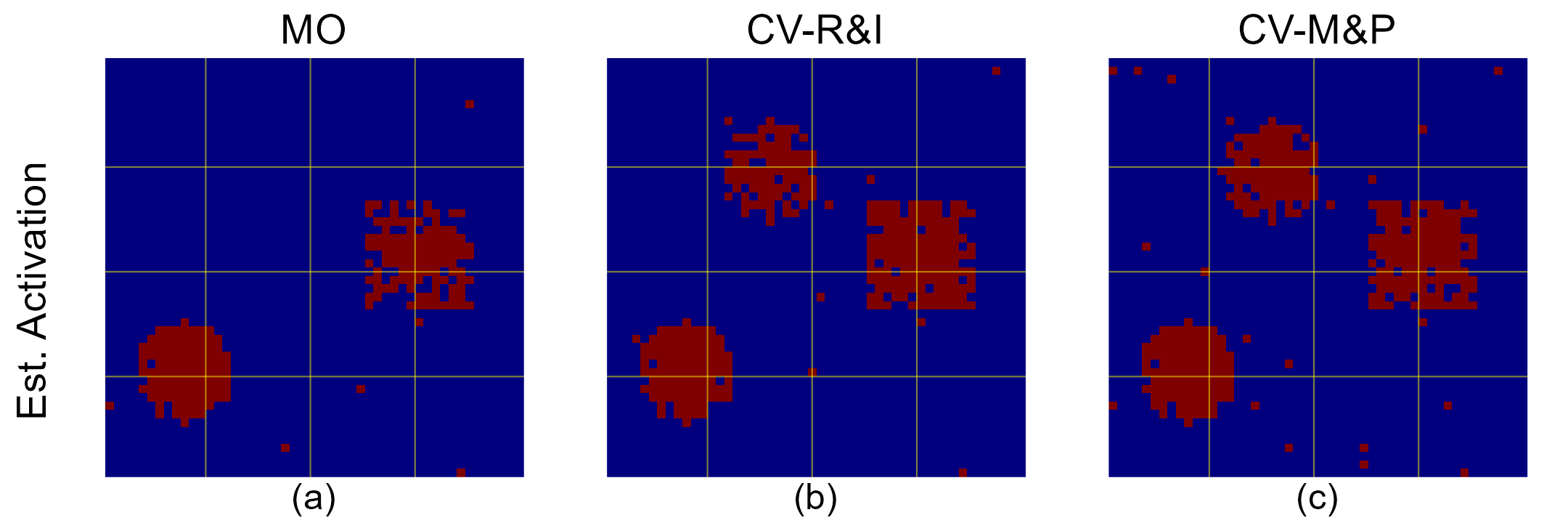}
    \end{center}
    \caption{(a)-(c) are estimated activation maps as derived from MO, CV-R\&I, and CV-M\&P, respectively.}
    \label{fig:MySim_comparison}
\end{figure}

By synthesizing the estimated activation maps for both magnitude and phase (Figures~\ref{fig:MySim_estimated_maps}b and \ref{fig:MySim_estimated_maps}f), we construct a composite activation map and compare it against results from MO and CV-R\&I. Figure~\ref{fig:MySim_comparison} presents these comparative maps. Performance evaluation reveals that MO fails the competition, primarily due to its inability to detect the phase-only active region 2. Conversely, both CV-R\&I and CV-M\&P deliver competitive results.

\begin{figure}
    \begin{center}
        \includegraphics[width=\textwidth]{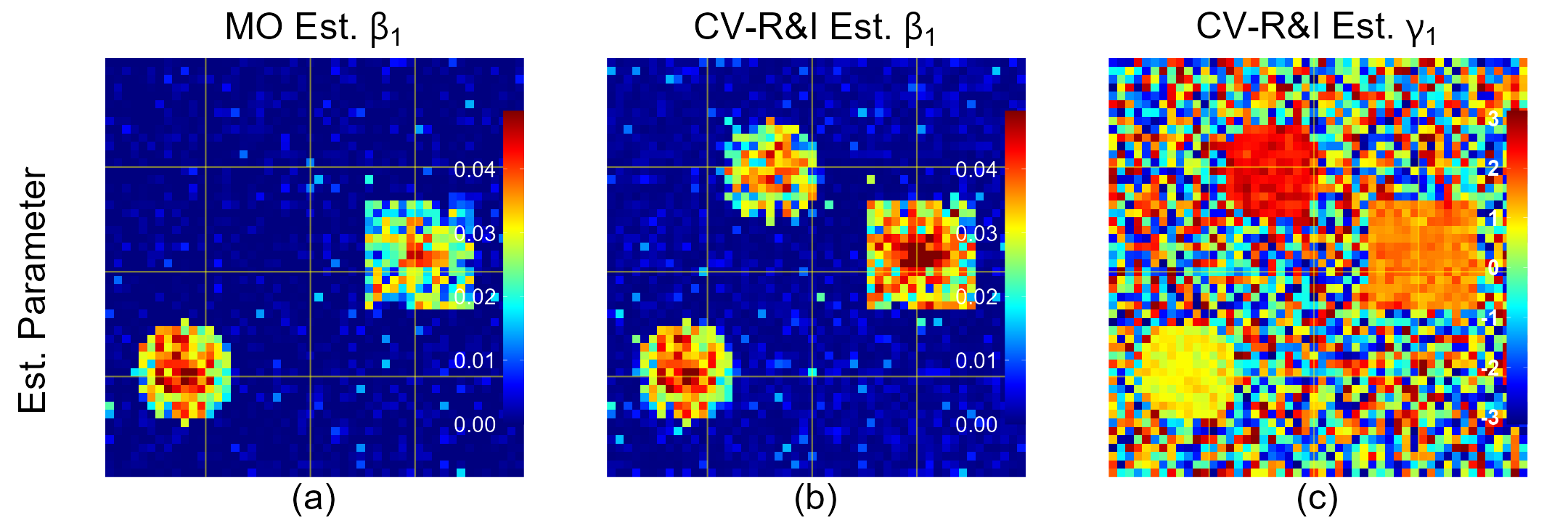}
    \end{center}
    \caption{(a)-(c) are improperly estimated parameters as derived from MO and CV-R\&I.}
    \label{fig:MySim_wrong_estimations}
\end{figure}

The analysis also extends to comparing the parameter estimations across the three models. As MO and CV-R\&I do not explicitly characterize parameters $\beta_1$ and $\gamma_1$ in their models, we resort to indirect methods for their estimation. For MO model, we use the estimated slope of the BOLD signal, $\widehat{\beta}_{v, M_1}$, as an estimate for $\beta_{v, 1}$, while for CV-R\&I, the square root of the sum of squares of the estimated slopes, $\sqrt{\left(\widehat{\beta}_{v, R_1}\right)^2+\left(\widehat{\beta}_{v, I_1}\right)^2}$, serves as an estimate for $\beta_{v, 1}$. As for $\gamma_1$, MO cannot estimate this parameter due to its limitation to magnitude-only data. In contrast, CV-R\&I employs $\text{arctan}_4\left(\widehat{\beta}_{v, I_1}/\widehat{\beta}_{v, R_1}\right)$ as an estimate for $\gamma_1$. These results are illustrated in Figure~\ref{fig:MySim_wrong_estimations}. Upon examination of Figure~\ref{fig:MySim_wrong_estimations}a, we observe that while MO’s estimated $\beta_1$ map appears to closely align with the true $\beta_1$ map (Figure~\ref{fig:MySim_estimated_maps}c), it still slightly underestimates values in region 3. Similarly, as seen in Figure~\ref{fig:MySim_wrong_estimations}b, CV-R\&I not only falsely estimates the non-existent $\beta_1$ in the phase-only active region 2, but also tends to overestimate $\beta_1$ in region 3. This overestimation of $\beta_1$ in region 3 where voxels exhibit both magnitude- and phase-active, is consistent with the findings in \cite{Wang2023}. Lastly, Figure~\ref{fig:MySim_wrong_estimations}c reveals the CV-R\&I's estimated $\gamma_1$ map significantly deviates from the true $\gamma_1$ map, as showcased in Figure~\ref{fig:MySim_estimated_maps}g.

The numerical evaluation metrics are summarized in Table~\ref{tab:MySim part1}, where we employ accuracy, precision, recall, F1-score, and the area under the receiver operating characteristic curve (ROC-AUC) to gauge classification performance. We also employ the regression slope between true and estimated parameters to quantify the estimation performance, to expect it be close to one. In terms of classification, CV-M\&P outperforms its counterparts in various key metrics, including recall, F1-score, and AUC. While the margin of superiority may not be pronounced, CV-M\&P offers two distinct advantages over its counterparts: it allows for the independent prediction of magnitude and phase activation maps, as shown in Figure~\ref{fig:MySim_estimated_maps}b and \ref{fig:MySim_estimated_maps}f, while the other two approaches cannot, and provides accurate estimation for both $\beta_1$ and $\gamma_1$, as the slopes (0.9731 and 0.9462) are close to one. Further evidence from multiple simulation studies, to be discussed in the subsequent section, will reinforce these findings.

\begin{table}
    \caption{Metrics of a single simulated dataset produced by the MO, CV-R\&I, and CV-M\&P models.}
    \footnotesize
    \begin{center}
        \begin{tabular}{c|cccccccc}
        Model & Accuracy & Precision & Recall & F1 Score & AUC & $\beta_1$ slope & $\gamma_1$ slope & Time (s)\\
        \hline
        MO & 0.9248 & 0.9726 & 0.5392 & 0.6938 & 0.8910 & 0.8630 & NA & \textbf{1.54} \\
        CV-R\&I & \textbf{0.9688} & \textbf{0.9731} & 0.8253 & 0.8931 & 0.9868 & 1.0301 & 33.427 & 3.93 \\
        CV-M\&P & 0.9680 & 0.9436 & \textbf{0.8481} & \textbf{0.8933} & \textbf{0.9896} & \textbf{0.9731} & \textbf{0.9462} & 17.87 \\
        \end{tabular}
    \end{center}
    \label{tab:MySim part1}
\end{table}


\subsection{Multiple Simulations}\label{sub: Multiple simulations}


\subsubsection{Generating random maps and simulating fMRI signals} We generate 100 random true strength maps using the parameters outlined in Table~\ref{tab:chapter3map parameters} and the \texttt{specifyregion} function. The true strength maps are then scaled by factors of 0.04909 and $\pi/36$ to obtain 100 true $\beta_{1}$ maps and 100 true $\gamma_{1}$ maps, respectively. Using Eq. \eqref{MySim_sim} and the expected BOLD response $\mx$ in Figure~\ref{fig:True_maps}b, we generate three datasets from each pair of true $\beta_{1}$ and $\gamma_{1}$ maps with the following assignments:
\begin{itemize}
\item $\beta_{1}$ present, $\gamma_{1}$ absent (all active voxels are solely magnitude-active)
\item $\beta_{1}$ absent, $\gamma_{1}$ present (all active voxels are solely phase-active)
\item $\beta_{1}$ present, $\gamma_{1}$ present (all active voxels are both magnitude- and phase-active)
\end{itemize}
The values for $\beta_{0}$, $\gamma_{0}$, and $\sigma$ are held constant as specified in Section \ref{subsection: Single simulation}, with values 0.4909, $\pi/4$, and 0.04909, respectively, resulting $\text{CNR}_\text{Mag}=1$ and $\text{CNR}_\text{Ph}=(\pi/36)/10$. In total, we have 300 datasets for analysis.

\begin{table}
    \caption{Characteristics of true maps.}
    \begin{center}
        \begin{tabular}{c|c|c|c|c}
        Map size & Number of active regions & Radius & Shape & Decay rate $(\varrho)$\\\hline
        50$\times$50 & 3 & 2 to 6 & sphere or cube & 0 to 0.3 \\
        \end{tabular}
    \end{center}
    \label{tab:chapter3map parameters}
\end{table}


\subsubsection{Results} Table \ref{tab:MySim multiple part1} presents the performance metrics for each method across these diverse datasets. In terms of classification,  MO delivers superior performance in almost all evaluated metrics for datasets featuring exclusively magnitude-active voxels, which is expected given its design specificity for magnitude-based activity. However, such an assumption of magnitude-only activity is often unrealistic in real-world applications. When considering datasets comprising solely phase-active voxels, CV-M\&P excels in all metrics except precision, thereby establishing its superiority in detecting phase-based activity. For mixed activity involving both magnitude and phase, CV-R\&I takes the lead in accuracy, precision, and F1-score metrics, whereas CV-M\&P dominates in recall and AUC.

\begin{table}
    \caption{Summary of average metrics across 100 simulated datasets produced by the MO, CV-R\&I, and CV-M\&P models. The values in parentheses are min, max, and standard deviation.}
    \centering
    \footnotesize
        \begin{tabular}{p{0.4cm}|c|c|c|c}
        Data Type & Measure & MO & CV-R\&I & CV-M\&P \\
        \hline
        \multirow{7}{*}{\shortstack{Mag. \\ -only}} & Accuracy & \textbf{0.9645}(0.9208, 0.9940, 0.0153) & 0.9555(0.9032, 0.9916, 0.0188) & 0.9598(0.9132, 0.9900, 0.0160) \\
        & Precision & \textbf{0.9647}(0.9157, 0.9917, 0.0164) & 0.9601(0.9078, 0.9964, 0.0166) & 0.9317(0.8515, 0.9805, 0.0224) \\
        & Recall & \textbf{0.7629}(0.6052, 0.9680, 0.0707) & 0.6966(0.5263, 0.9406, 0.0840) & 0.7534(0.6111, 0.9634, 0.0721) \\
        & F1 Score &  \textbf{0.8502}(0.7366, 0.9716, 0.0437) & 0.8046(0.6741, 0.9515, 0.0557) & 0.8311(0.7294, 0.9444, 0.0435)\\
        & AUC &  0.9760(0.9485, 0.9991, 0.0107) & 0.9605(0.9227, 0.9963, 0.0154) & \textbf{0.9793}(0.9605, 0.9983, 0.0081)\\
        & $\beta_1$ slope &  0.8696(0.7927, 0.9466, 0.0327) & 0.8337(0.7451, 0.9356, 0.0406) & \textbf{0.9771}(0.9337, 1.0170, 0.0190)\\
        & $\gamma_1$ slope &  NA & NA & NA\\
        \hline
        \multirow{7}{*}{\shortstack{Ph. \\ -only}} & Accuracy & 0.8638(0.7576, 0.9428, 0.0418) & 0.9390(0.8696, 0.9848, 0.0234) & \textbf{0.9459}(0.8868, 0.9832, 0.0201) \\
        & Precision & 0.2102(0.0714, 0.6000, 0.1097) & \textbf{0.9481}(0.8829, 0.9862, 0.0207) & 0.9192(0.8324, 0.9735, 0.0303) \\
        & Recall & 0.0059(0.0017, 0.0201, 0.0034) & 0.5718(0.4207, 0.9026, 0.0970) & \textbf{0.6481}(0.5146, 0.9090, 0.0881) \\
        & F1 Score &  0.0114(0.0034, 0.0373, 0.0065) & 0.7088(0.5831, 0.9269, 0.0725) & \textbf{0.7569}(0.6456, 0.9225, 0.0604)\\
        & AUC &  0.5277(0.4898, 0.6001, 0.0214) & 0.9326(0.8844, 0.9930, 0.0237) & \textbf{0.9544}(0.9216, 0.9952, 0.0150)\\
        & $\beta_1$ slope &  NA & NA & NA\\
        & $\gamma_1$ slope &  NA & 39.813(30.031, 45.799, 3.3285) & \textbf{0.9439}(0.8744, 1.0271, 0.0289)\\
        \hline
        \multirow{7}{*}{Both} & Accuracy & 0.9644(0.9192, 0.9912, 0.0144) & \textbf{0.9835}(0.9616, 0.9984, 0.0071) & 0.9769(0.9544, 0.9896, 0.0069) \\
        & Precision & 0.9643(0.9017, 0.9892, 0.0146) & \textbf{0.9798}(0.9354, 0.9967, 0.0106) & 0.9134(0.7870, 0.9617, 0.0294) \\
        & Recall & 0.7606(0.6358, 0.9662, 0.0651) & 0.8949(0.8216, 0.9925, 0.0362) & \textbf{0.9073}(0.8457, 0.9927, 0.0329) \\
        & F1 Score &  0.8489(0.7703, 0.9592, 0.0398) & \textbf{0.9350}(0.8926, 0.9908, 0.0202) & 0.9097(0.8299, 0.9539, 0.0208)\\
        & AUC &  0.9763(0.9537, 0.9992, 0.0103) & 0.9939(0.9846, 0.9999, 0.0035) & \textbf{0.9940}(0.9873, 0.9997, 0.0026)\\
        & $\beta_1$ slope &  0.8710(0.7879, 0.9789, 0.0321) & 1.2346(1.1693, 1.3360, 0.0305) & \textbf{0.9843}(0.9365, 1.0316, 0.0183)\\
        & $\gamma_1$ slope &  NA & 26.643(19.307, 30.993, 2.4747) & \textbf{0.9534}(0.8958, 1.0146, 0.0253)\\
        \end{tabular}

    \label{tab:MySim multiple part1}
\end{table}

CV-M\&P once again stands out with respect to parameter estimation. Specifically, its true vs estimated parameter slopes are close to one when the parameters are present in the simulation, indicating accurate estimations. In contrast, this metric from both MO and CV-R\&I deviates from the ideal value of one. As explained by \cite{Wang2023}, under conditions where all active voxels are solely magnitude-active, the Cartesian model of \cite{Lee2007} and the polar model of  \cite{Rowe2005c} (CV-R\&I and CV-M\&P, respectively) are approximately equivalent. Hence, in such scenario, CV-R\&I can properly estimate $\beta_1$, although not surpassing the performance of CV-M\&P. In other scenarios, both MO and CV-R\&I fall short, either failing to estimate $\gamma_1$ or inaccurately estimating both $\beta_1$ and $\gamma_1$.


\section{Analysis of Human CV-fMRI Data}\label{RealStudy}
In this study, we employ the experimental data previously analyzed by \cite{Yu2018, Yu2023} and \cite{Wang2023}. This dataset originates from a unilateral finger-tapping experiment conducted using a 3.0-Tesla General Electric Signa LX MRI scanner. The experimental design comprises 16.33 epochs, each consisting of alternating periods of 15s on and 15s off. Consequently, the total number of time points is $T=490$, excluding the warm-up phase. The acquired dataset has seven slices, each with dimensions $96\times 96$, and our analysis focuses on the initial six slices. For all examined models, we set the number of parcels $G=25$. Specific tuning parameters are fixed based on the experience: for the MO model, $\psi=\Phi^{-1}(0.02)$; for CV-R\&I, $\psi=\Phi^{-1}(0.1)$; and for CV-M\&P, $\psi=\omega=\Phi^{-1}(0.20)$. The thresholds for identifying active voxels are set to values as follows: 0.8722 for both MO and CV-R\&I, and 0.925 for CV-M\&P, in alignment with Section~\ref{sbusec:MCMC} and the simulation studies.

\begin{figure}
    \begin{center}
    \includegraphics[width=\textwidth]{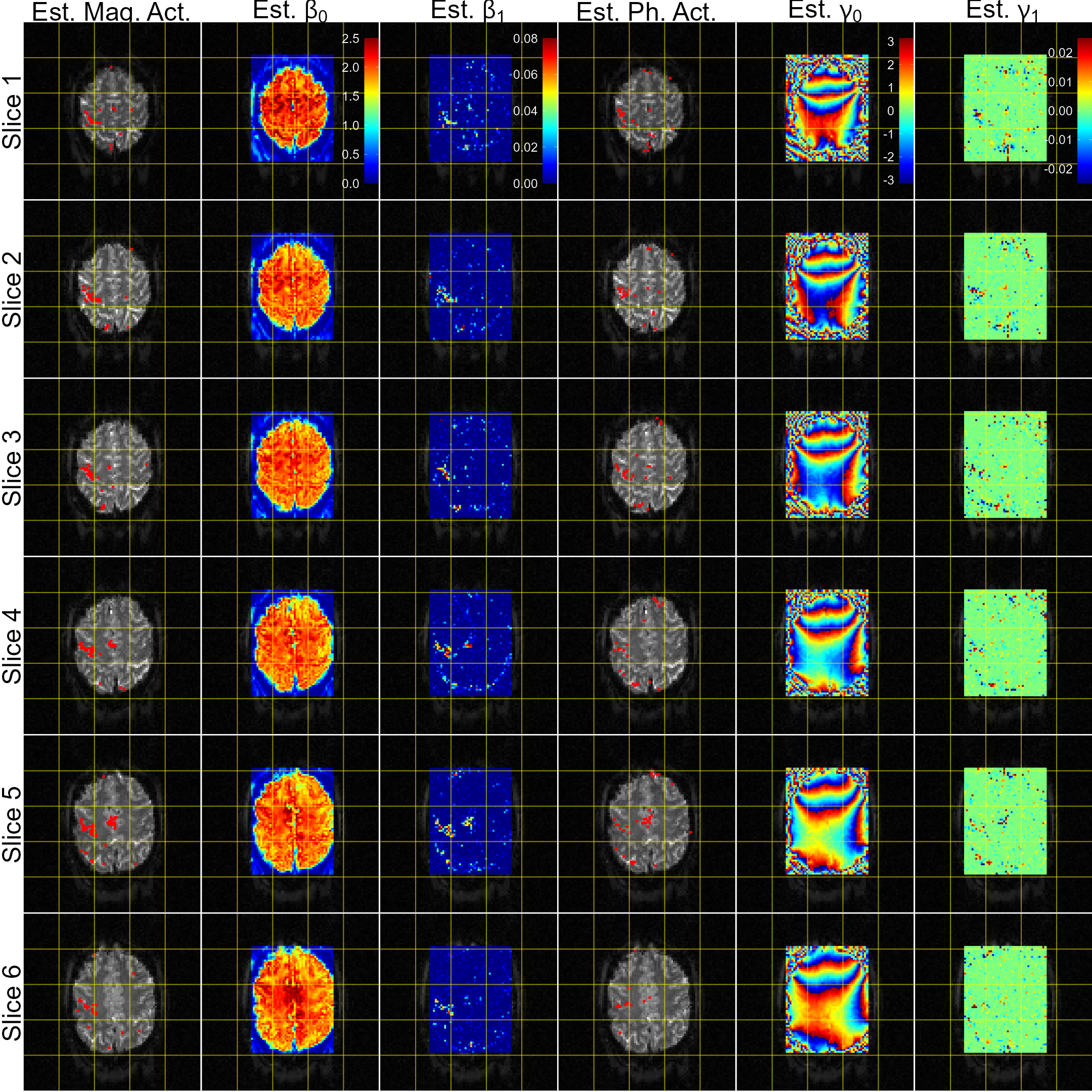}
    \end{center}
    \caption{Estimated magnitude activation, $\beta_0$, $\beta_1$, phase activation, $\gamma_1$, $\gamma_0$ maps for a real human brain dataset as derived by the CV-M\&P model.}
    \label{fig:RealHuman_estimated_maps}
\end{figure}

In Figure~\ref{fig:RealHuman_estimated_maps}, we present the results derived from the CV-M\&P model. Distinct patterns are observed: the estimated $\beta_0$ maps mirror the patterns of magnitude in the background, the estimated $\gamma_0$ maps highlight the phase's transition lines across different color zones, and both the estimated $\beta_1$ and $\gamma_1$ maps reflect patterns consistent with the estimated magnitude and phase activation maps. Such patterns are indicative of the accuracy of our approach in both classification and estimation.

\begin{figure}
    \begin{center}
    \includegraphics[width=\textwidth]{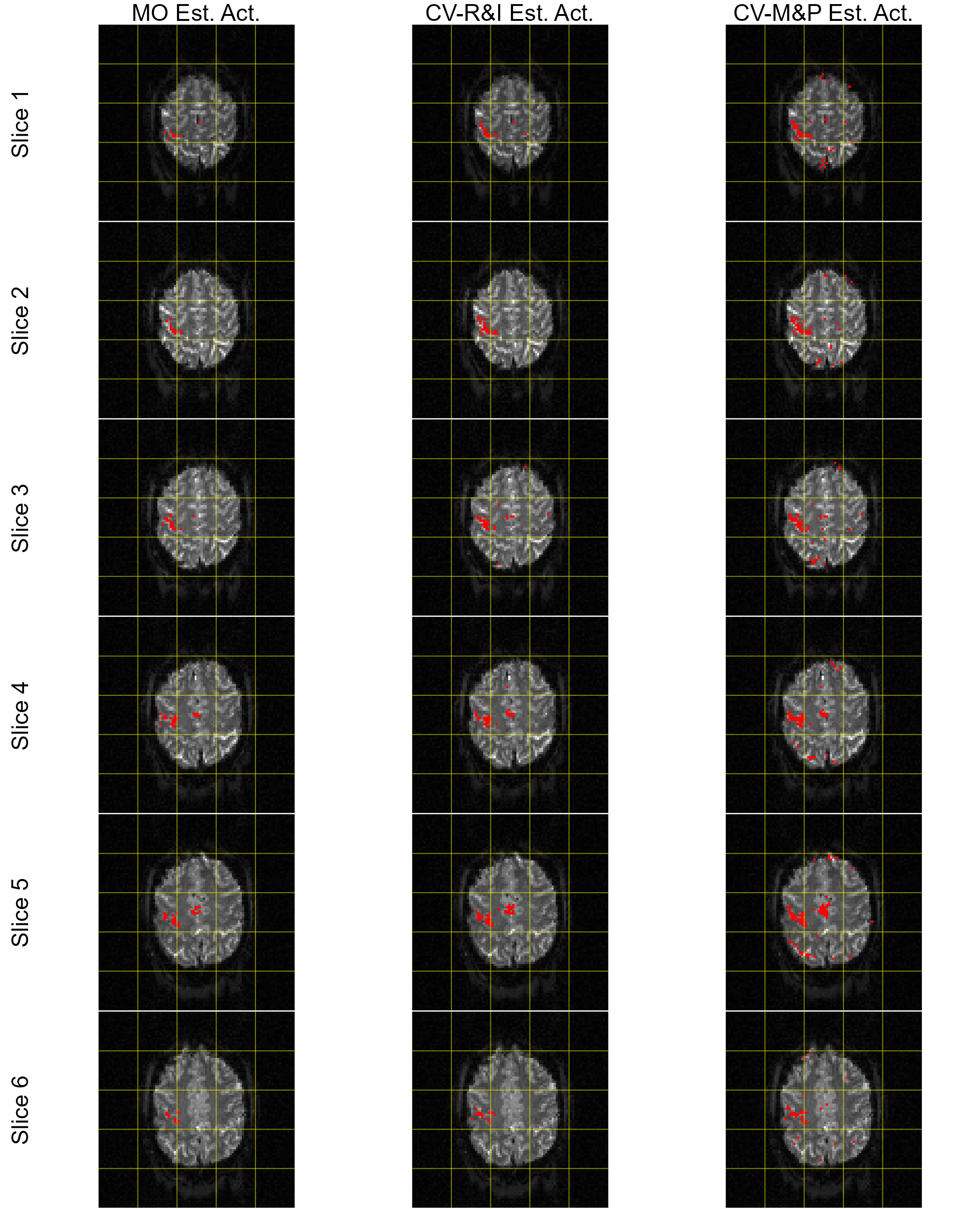}
    \end{center}
    \caption{Estimated activation maps for a real human brain dataset as derived by the MO, CV-R\&I, and CV-M\&P models.}
    \label{fig:RealHuman_comparison}
\end{figure}

By integrating the magnitude- and phase-activation maps derived by CV-M\&P model, we form comprehensive estimated activation maps. They are subsequently compared with activation maps estimated by MO and CV-R\&I models, as shown in Figure~\ref{fig:RealHuman_comparison}, revealing significant alignment. Specifically, the two central and central-left active regions detected by CV-M\&P are consistent with the findings reported in \cite{Yu2018, Yu2023} and \cite{Wang2023}. Furthermore, these regions align with known anatomical areas typically activated during finger-tapping tasks. The central region may correspond to the Primary Motor Cortex (M1) or Supplementary Motor Area (SMA), both of which play pivotal roles in voluntary movement and motor planning \citep{Penfield1937, Geyer1996}. Adjacently, the central-left region might represent the Primary Somatosensory Cortex (S1) or the Posterior Parietal Cortex, responsible for tactile sensory information processing and sensory-motor integration, respectively \citep{Culham2006}. Notably, beyond these well-established regions, CV-M\&P uncovers additional active regions at the posterior of the brain image. These could be caused by the motion of brain during the data collection.

In terms of computational efficiency, MO, CV-R\&I, and CV-M\&P required 12.32 seconds, 32.59 seconds, and 910.2 seconds, respectively, to complete $10^3$ iterations. Despite CV-M\&P's computational demand and increased execution time, its drawbacks are mitigated by its potential for scalability. As highlighted in the works of \cite{Musgrove2016} and \cite{Wang2023}, the employment of a brain parcellation strategy along with the sSGLMM spatial prior has minimal effects on the prediction results, as long as the total parcel number $G$ is in a reasonable range. Given that this dataset can be further divided into more parcels for parallel processing, and considering our computational resources are currently limited to 16 CPU cores, the computational efficiency of the CV-M\&P model can be substantially improved.


\section{Conclusion}\label{Conclusion}
Throughout our investigations on both simulated and real human datasets, the CV-M\&P model consistently demonstrates its capability to precisely identify voxels that exhibit significant reactions to stimuli, whether in magnitude, in phase, or in a combination of both. Comparing with the polar model of \cite{Rowe2005c}, but using hypothesis testing approaches \citep{Rowe2004, Rowe2005a, Rowe2005c, Rowe2005b, Rowe2007, Rowe2009a, Adrian2018}, our fully Bayesian framework can capture the spatial correlations of fMRI data, and therefore improve the model flexibility. On the other hand, comparing with other fully Bayesian approaches, but based on the Cartesian model of \cite{Lee2007} \citep{Yu2023, Wang2023}, our CV-M\&P model rectifies the constraints inherent in the Cartesian models, which can detect active voxels but remains ambiguous about the exact type of the activation \citep{Rowe2009a}. Moreover, the CV-M\&P model excels in providing precise parameter estimates, offering a more nuanced framework for delineating brain activation patterns in task-based fMRI analyses. 

There are multiple avenues for advancing this research. These include the exploration of more complex models that account for temporal correlations, models that fit the non-circular data wherein the real and imaginary components of the signal are correlated, and efforts aimed at optimizing computational efficiency.


\section*{Acknowledgements}
This research is supported by the National Institute of General Medical Sciences of the National Institutes of Health under award number P20GM139769 (X. Li), and National Science Foundation awards DMS-2210658 (X. Li) and DMS-2210686 (D. A. Brown). The content is solely the responsibility of the authors and does not necessarily represent the official views of the National Institutes of Health or the National Science Foundation.

\clearpage
\bibliographystyle{plainnat}
\bibliography{sample}

\newpage
\appendix

\begin{center}
\huge Appendix
\end{center}


\section{Conditional Posterior Distributions for Gibbs Sampling}
We need the full conditional posterior distributions of
\begin{equation}
    \mbeta_v, \lambda_v, \mgamma_v, \omega_v, \sigma_v^2, \tau_g^2, \xi_g^2, \eta_{\lambda, v}, \mdelta_{\lambda, g}, \kappa_{\lambda, g}, \eta_{\omega, v}, \mdelta_{\omega, g}, \kappa_{\omega, g}
\end{equation}
for the Gibbs sampling. All derivations will omit the subscript of $g$ (parcel index) from the parcel-level parameters $\tau_g^2, \xi_g^2, \mdelta_{\lambda, g}, \kappa_{\lambda, g}, \mdelta_{\omega, g}, \kappa_{\omega, g}$, since all parcels run the algorithm identically.


\subsection{Full conditional distributions of $\lambda_v$ and $\omega_v$}
The full conditional distribution of $\lambda_v$ is
\begin{equation}
    \pi(\lambda_v\mid\my_r^v, \mbeta_v, \mgamma_v, \omega_v, \sigma_v^2, \tau^2, \xi^2, \eta_{\lambda, v})={\B}ern\left(P_{\lambda_v}\right),
\end{equation}
where
\begin{equation}
    \begin{split}
        P_{\lambda_v}&=p(\lambda_v=1\mid\my_r^v, \mbeta_v, \mgamma_v, \omega_v, \sigma_v^2, \tau^2, \xi^2, \eta_{\lambda, v})\\
        &=\frac{p(\lambda_v=1\mid\eta_{\lambda, v})}{p(\lambda_v=1\mid\eta_{\lambda, v})+\frac{L_0}{L_1}{\cdot}p(\lambda_v=0\mid\eta_{\lambda, v})}\\
        &=\frac{\mPhi(\psi_{\lambda}+\eta_{\lambda, v})}{\mPhi(\psi_{\lambda}+\eta_{\lambda, v})+\frac{L_0}{L_1}{\cdot}\left[1-\mPhi(\psi_{\lambda}+\eta_{\lambda, v})\right]},
    \end{split}
\end{equation}
and $L_0$ and $L_1$ are the joint densities of $\my_v, \mbeta_v, \lambda_v, \mgamma_v, \omega_v, \sigma_v^2, \tau^2, \xi^2$ given $\lambda_v=0$ and $\lambda_v=1$. Let $L$ be such joint density, that is,
\begin{equation}
    L=p(\my_v, \mbeta_v, \lambda_v, \mgamma_v, \omega_v, \sigma_v^2, \tau^2, \xi^2)\propto p(\my_v\mid\mbeta_v, \lambda_v, \mgamma_v, \omega_v, \sigma_v^2)p(\mbeta_v\mid\lambda_v, \tau^2).
\end{equation}
Define:
\begin{equation}
    \mA_v=\begin{pmatrix}
            \mC_v\\
            \mS_v\\
         \end{pmatrix}
    , \quad \text{where }\mC_v=\text{diag}\left[\cos{\left(\mU\mOmega_v\mgamma_v\right)}\right], \quad \mS_v=\text{diag}\left[\sin{\left(\mU\mOmega_v\mgamma_v\right)}\right],\\
\end{equation}
(Note, $\mA_v$ is orthogonal, i.e., $\mA_v'\mA_v=\mI_{2T}$.), then,
\begin{equation}
    \begin{split}
        &\quad p(\my_v\mid\mbeta_v, \lambda_v, \mgamma_v, \omega_v, \sigma_v^2)\\
        &=(2\pi\sigma_v^2)^{-\frac{2T}{2}}\exp{\left\{-\frac{1}{2\sigma_v^2}\left(\my_v-\mA_v\mX
        \mLambda_v\mbeta_v\right)'\left(\my_v-\mA_v\mX
        \mLambda_v\mbeta_v\right)\right\}}\\
        &=(2\pi\sigma_v^2)^{-\frac{2T}{2}}\exp{\left\{-\frac{1}{2\sigma_v^2}\left[\my_v'\my_v-2\left(\mA_v\mX        \mLambda_v\mbeta_v\right)'\my_v+\left(\mLambda_v\mbeta_v\right)'\mX'\mX\mLambda_v\mbeta_v\right]\right\}},
    \end{split}
\end{equation}
and
\begin{equation}
    p(\mbeta_v\mid \lambda_v, \tau^2)=(2\pi\tau^2)^{-\frac{1+\lambda_v}{2}}\exp{\left\{-\frac{1}{2\tau^2}\mbeta_v'\mLambda_v\mbeta_v\right\}}.
\end{equation}
Thus,
\begin{equation}
    L\propto (2\pi\tau^2)^{-\frac{1+\lambda_v}{2}}\exp{\left\{-\frac{1}{2\sigma_v^2}\left[-2\left(\mA_v\mX        \mLambda_v\mbeta_v\right)'\my_v+\left(\mLambda_v\mbeta_v\right)'\mX'\mX\mLambda_v\mbeta_v\right]-\frac{1}{2\tau^2}\mbeta_v'\mLambda_v\mbeta_v\right\}}.
\end{equation}
Let $\ma_v$ be the flattened version of $\mA_v$, that is, $\ma_v$ is a $2T\times 1$ vector as
\begin{equation}
    \ma_v=\mA_v\mone_{T}=\begin{pmatrix}
    \cos{\left(\mU\mOmega_v\mgamma_v\right)}\\
    \sin{\left(\mU\mOmega_v\mgamma_v\right)}
\end{pmatrix}
\end{equation}
Also, define $\mx_{(2)}$ as the second column of $\mX$, thus, $\mx_{(2)}$ is a $T\times 1$ vector of expected BOLD response; define $\mx_{(2)}^*=\begin{pmatrix}
    \mx_{(2)}\\
    \mx_{(2)}
\end{pmatrix}$ as a $2T\times 1$ vector to match the dimension, then,
\begin{equation}
    \begin{split}
        \frac{L_0}{L_1}&=\frac{L\mid_{\lambda_v=0}}{L\mid_{\lambda_v=1}}\\
        &=(2\pi\tau^2)^{\frac{1}{2}}\exp{\left\{-\frac{1}{2\sigma_v^2}\left[2\beta_{v, 1}\left(\mx_{(2)}^*\odot\ma_v\right)'\my_v-2\beta_{v, 0}\beta_{v, 1}\mx_{(2)}'\mone_{T}-\beta_{v, 1}^2\mx_{(2)}'\mx_{(2)}\right]+\frac{1}{2\tau^2}\beta_{v, 1}^2\right\}}.
    \end{split}
\end{equation}
We flatten $\mA_v$ and use Hadamard product $\odot$ here to lessen the computational burden. Similarly, the full conditional distribution of $\omega_v$ is
\begin{equation}
    \pi(\omega_v\mid\my_r^v, \mbeta_v, \lambda_v, \mgamma_v, \sigma_v^2, \tau^2, \xi^2, \eta_v)={\B}ern\left(P_{\omega_v}\right),
\end{equation}
where
\begin{equation}
    P_{\omega_v}=\frac{\mPhi(\psi_{\omega}+\eta_{\omega, v})}{\mPhi(\psi_{\omega}+\eta_{\omega, v})+\frac{L_0}{L_1}{\cdot}\left[1-\mPhi(\psi_{\omega}+\eta_{\omega, v})\right]},
\end{equation}
where
\begin{equation}
    L\propto (2\pi\xi^2)^{-\frac{1+\omega_v}{2}}\exp{\left\{\frac{1}{\sigma_v^2}\left(\mA_v\mX\mLambda_v\mbeta_v\right)'\my_v-\frac{1}{2\xi^2}\mgamma_v'\mOmega_v\mgamma_v\right\}},
\end{equation}
and
\begin{equation}
    \begin{split}
        \frac{L_0}{L_1}&=\frac{L\mid_{\omega_v=0}}{L\mid_{\omega_v=1}}\\
        &=(2\pi\xi^2)^{\frac{1}{2}}\exp{\left\{\frac{1}{\sigma_v^2}\left[\left(\mA_v\mid_{\omega_v=0}-\mA_v\mid_{\omega_v=1}\right)\mX\mLambda_v\mbeta_v\right]'\my_v+\frac{1}{2\xi^2}\gamma_{v, 1}^2\right\}}.
    \end{split}
\end{equation}
Keep simplifying it, when $\lambda_v=0$,
\begin{equation}
    \frac{L_0}{L_1}=(2\pi\xi^2)^{\frac{1}{2}}\exp{\left\{\frac{1}{\sigma_v^2}\beta_{v, 0}\left(\ma_v\mid_{\omega_v=0}-\ma_v\mid_{\omega_v=1}\right)'\my_v+\frac{1}{2\xi^2}\gamma_{v, 1}^2\right\}}.
\end{equation}
When $\lambda_v=1$,
\begin{equation}
    \frac{L_0}{L_1}=(2\pi\xi^2)^{\frac{1}{2}}\exp{\left\{\frac{1}{\sigma_v^2}\left[\left(\beta_{v, 0}\mone_{2T}+\beta_{v, 1}\mx_{(2)}^*\right)\odot\left(\ma_v\mid_{\omega_v=0}-\ma_v\mid_{\omega_v=1}\right)\right]'\my_v+\frac{1}{2\xi^2}\gamma_{v, 1}^2\right\}}.
\end{equation}


\subsection{Full conditional distribution of $\mbeta_v$} When $\lambda_v=1$, the full conditional distribution of $\mbeta_v$ is
\begin{equation}
    \begin{split}
        &\quad \pi(\mbeta_v\mid \my_v, \lambda_v=1, \mgamma_v, \omega_v, \sigma_v^2, \tau^2, \xi^2)\\
        &\propto p(\my_v, \mbeta_v, \lambda_v=1, \mgamma_v, \omega_v, \sigma_v^2, \tau^2, \xi^2)\\
        &\propto p(\my_v\mid\mbeta_v, \lambda_v=1, \mgamma_v, \omega_v, \sigma_v^2)p(\mbeta_v\mid\lambda_v=1, \tau^2)\\
        &\propto \exp{\left\{-\frac{1}{2\sigma_v^2}\left(\my_v-\mA_v\mX
        \mbeta_v\right)'\left(\my_v-\mA_v\mX
        \mbeta_v\right)\right\}}
        \exp{\left\{-\frac{1}{2\tau^2}\mbeta_v'\mbeta_v\right\}}\\
        &\propto \exp{\left\{ -\frac{1}{2}\left[\mbeta_v'\frac{\left(\mA_v\mX
        \right)'\left(\mA_v\mX
        \right)}{\sigma_v^2}\mbeta_v-2\mbeta_v'\frac{\left(\mA_v\mX
        \right)'}{\sigma_v^2}\my_v+\mbeta_v'\frac{1}{\tau^2}\mbeta_v\right] \right\}}\\
        &=\exp{\left\{ -\frac{1}{2}\left[\mbeta_v'\frac{\left(\mA_v\mX
        \right)'\left(\mA_v\mX
        \right)+\frac{\sigma_v^2}{\tau^2}\mI}{\sigma_v^2}\mbeta_v-2\mbeta_v'\frac{\left(\mA_v\mX
        \right)'}{\sigma_v^2}\my_v\right] \right\}}\\
        &=\exp{\left\{ -\frac{1}{2}\left[\mbeta_v'\frac{\mX'\mX
        +\frac{\sigma_v^2}{\tau^2}\mI}{\sigma_v^2}\mbeta_v-2\mbeta_v'\frac{\left(\mA_v\mX
        \right)'}{\sigma_v^2}\my_v\right] \right\}}.
    \end{split}
\end{equation}
Therefore,
\begin{equation}
    \pi(\mbeta_v\mid \my_v, \lambda_v=1, \mgamma_v, \omega_v, \sigma_v^2, \tau^2, \xi^2)=\N_{2}(\mmu_{\mbeta_v}, \mSigma_{\mbeta_v}),
\end{equation}
where
\begin{equation}
    \begin{split}
        \mmu_{\mbeta_v}&=\left(\mX'\mX
        +\frac{\sigma_v^2}{\tau^2}\mI\right)^{-1}\left(\mA_v\mX
        \right)'\my_v,\\
        \mSigma_{\mbeta_v}&=\sigma_v^2\left(\mX'\mX
        +\frac{\sigma_v^2}{\tau^2}\mI\right)^{-1},
    \end{split}
\end{equation}
where $\mA_v\mX$ can be calculated as $\left[\ma_v, \mx_{(2)}^*\odot\ma_v\right]$ for faster computation. When $\lambda_v=0$, it's easy to show:
\begin{equation}
    \pi(\mbeta_{v, 0}\mid \my_v, \lambda_v=0, \mgamma_v, \omega_v, \sigma_v^2, \tau^2, \xi^2)=\N(\frac{\left(\mA_v\mone_{T}\right)'\my_v}{T+\frac{\sigma_v^2}{\tau^2}}, \frac{\sigma_v^2}{T+\frac{\sigma_v^2}{\tau^2}}),
\end{equation}
and $\mbeta_{v, 1}=0$ with probability 1, where $\mA_v\mone_{T}$ is just $\ma_v$.


\subsection{Sampling $\mgamma_v$}
We apply Metropolis-Hastings algorithm to sample $\mgamma_v$. A random walk proposal,
\begin{equation}
    \mgamma_v^*\mid \mgamma_v\sim\N_2\left(\mOmega_v\mgamma_v, \quad \mOmega_v'\mSigma_{\mgamma_v}\mOmega_v\right),
\end{equation}
is used, where $\mgamma_v^*$ and $\mgamma_v$ are proposed parameter and current state, respectively, and $\mSigma_{\mgamma_v}$ is a tuning parameter. We use the current indicator of phase status, $\mOmega_v$, to secure it proposes $\mgamma_v^*=\begin{pmatrix}
    \gamma_{v, 0}^*\neq 0\\
    \gamma_{v, 1}^*=0
\end{pmatrix}$ when the phase is inactive. Let $p_{\mgamma_v}(\cdot)$ be the proposal density, then the acceptance ratio is
\begin{equation}
    \begin{split}
        r_{\mgamma_v}&=\frac{\pi(\mgamma_v^*\mid \my_v, \mbeta_v, \lambda_v, \omega_v, \sigma_v^2, \tau^2, \xi^2)p_{\mgamma_v}(\mgamma_v\mid \mgamma_v^*)}{\pi(\mgamma_v\mid \my_v, \mbeta_v, \lambda_v, \omega_v, \sigma_v^2, \tau^2, \xi^2)p_{\mgamma_v}(\mgamma_v^*\mid \mgamma_v)}\\
        &=\frac{p(\my_v\mid\mbeta_v, \lambda_v, \mgamma_v^*, \omega_v, \sigma_v^2)p(\mgamma_v^*\mid\omega_v, \xi^2)}{p(\my_v\mid\mbeta_v, \lambda_v, \mgamma_v, \omega_v, \sigma_v^2)p(\mgamma_v\mid\omega_v, \xi^2)},
    \end{split}
\end{equation}
where
\begin{equation}
    \begin{split}
        p(\my_v\mid\mbeta_v, \lambda_v, \mgamma_v, \omega_v, \sigma_v^2)&\propto \exp{\left\{-\frac{1}{2\sigma_v^2}\left(\my_v-\mA_v\mX
        \mLambda_v\mbeta_v\right)'\left(\my_v-\mA_v\mX
        \mLambda_v\mbeta_v\right)\right\}}\\
        &\propto\exp{\left\{\frac{1}{\sigma_v^2}\left(\mA_v\mX\mLambda_v\mbeta_v\right)'\my_v\right\}},\\
        p(\mgamma_v\mid\omega_v, \xi^2)&\propto \exp{\left\{-\frac{1}{2\xi^2}\mgamma_v'\mOmega_v\mgamma_v\right\}}.
    \end{split}
\end{equation}
Simplify the ratio, when $\lambda_v=0$,
\begin{equation}
    r_{\mgamma_v}=\exp{\left\{\frac{1}{\sigma_v^2}\beta_{v, 0}\left(\ma_v\mid_{\mgamma_v=\mgamma_v^*}-\ma_v\mid_{\mgamma_v=\mgamma_v}\right)'\my_v-\frac{1}{2\xi^2}\left(\gamma_{v, 0}^{*, 2}-\gamma_{v, 0}^2\right)\right\}}.
\end{equation}
When $\lambda_v=1$,
\begin{equation}
    r_{\mgamma_v}=\exp{\left\{\frac{1}{\sigma_v^2}\left[\left(\beta_{v, 0}\mone_{2T}+\beta_{v, 1}\mx_{(2)}^*\right)\odot\left(\ma_v\mid_{\mgamma_v=\mgamma_v^*}-\ma_v\mid_{\mgamma_v=\mgamma_v}\right)\right]'\my_v-\frac{1}{2\xi^2}\left(\mgamma_v^{*'}\mgamma_v^*-\mgamma_v'\mgamma_v\right)\right\}}.
\end{equation}
We generate a dummy variable $d_{\mgamma_v}\sim \U(0, 1)$, and if $d_{\mgamma_v}<r_{\mgamma_v}$, we update $\mgamma_v$ by $\mgamma_v^*$, otherwise remain $\mgamma_v$.


\subsection{Full conditional distribution of $\sigma_v^2$}
Assigning a Jeffreys prior, $p(\sigma_v^2)\propto 1/\sigma_v^2$, we have:
\begin{equation}
\pi(\sigma_v^2\mid\my_v, \cdot)=\I\G\left(\frac{2T}{2}, \quad\frac{1}{2}\left(\my_v-\mA_v\mX
        \mLambda_v\mbeta_v\right)'\left(\my_v-\mA_v\mX
        \mLambda_v\mbeta_v\right)\right).
\end{equation}
Again, to save computational time, $\mA_v\mX\mLambda_v\mbeta_v$ can be calculated as $\beta_{v, 0}\ma_v$ when $\lambda_v=0$, or $\left(\beta_{v, 0}\mone_{2T}+\beta_{v, 1}\mx_{(2)}^*\right)\odot\ma_v$ when $\lambda_v=1$.


\subsection{Full conditional distributions of $\tau^2$ and $\xi^2$}
The full conditional distribution of $\tau^2$ should be related to all voxels' $\beta_0$'s and magnitude-active voxels' $\beta_1$'s. Assigning a Jeffreys prior, $p(\tau^2)\propto 1/\tau^2$, we have:
\begin{equation}
\pi(\tau^2\mid\my_v, \cdot)=\I\G\left(\frac{1}{2}\sum_{v=1}^V \mone_2'\mLambda_v\mone_2, \quad\frac{1}{2}\sum_{v=1}^V\mbeta_v'\mLambda_v\mbeta_v\right).
\end{equation}
Equivalently,
\begin{equation}
\pi(\tau^2\mid\my_v, \cdot)=\I\G\left(\frac{1}{2}\left(V+\sum_{v=1}^V \lambda_v\right), \quad\frac{1}{2}\sum_{v=1}^V\left[\beta_{v, 0}^2+\left(\lambda_v\beta_{v, 1}\right)^2\right]\right).
\end{equation}
Similarly, the full conditional distribution of $\xi^2$ should be related to all voxels' $\gamma_0$'s and phase-active voxels' $\gamma_1$'s, that is,
\begin{equation}
\pi(\xi^2\mid\my_v, \cdot)=\I\G\left(\frac{1}{2}\sum_{v=1}^V \mone_2'\mOmega_v\mone_2, \quad\frac{1}{2}\sum_{v=1}^V\mgamma_v'\mOmega_v\mgamma_v\right).
\end{equation}
Equivalently,
\begin{equation}
\pi(\xi^2\mid\my_v, \cdot)=\I\G\left(\frac{1}{2}\left(V+\sum_{v=1}^V \omega_v\right), \quad\frac{1}{2}\sum_{v=1}^V\left[\gamma_{v, 0}^2+\left(\omega_v\gamma_{v, 1}\right)^2\right]\right).
\end{equation}


\subsection{Full conditional distributions of $\eta_v$, $\mdelta$, and $\kappa$} Let $\mQ_{s}=\mM'\mQ\mM$. then we follow the supplementary material in \cite{Wang2023}, we have:

\begin{equation}
\pi(\eta_{\lambda, v}\mid\lambda_v ,\kappa_{\lambda})= \begin{cases} \T\N(0, \; \frac{1}{\kappa_{\lambda}}\left(1+\mm_v{'}\mQ_{s}^{-1}\mm_v\right),\; 0,\; \infty) & { {\rm if }\; \lambda_v=1}\\ \T\N(0, \; \frac{1}{\kappa_{\lambda}}\left(1+\mm_v{'}\mQ_{s}^{-1}\mm_v\right),\; -\infty,\; 0) & { {\rm if }\; \lambda_v=0} \end{cases},
\end{equation}
\begin{equation}
\pi(\eta_{\omega, v}\mid\omega_v ,\kappa_{\omega})= \begin{cases} \T\N(0, \; \frac{1}{\kappa_{\omega}}\left(1+\mm_v{'}\mQ_{s}^{-1}\mm_v\right),\; 0,\; \infty) & { {\rm if }\; \omega_v=1}\\ \T\N(0, \; \frac{1}{\kappa_{\omega}}\left(1+\mm_v{'}\mQ_{s}^{-1}\mm_v\right),\; -\infty,\; 0) & { {\rm if }\; \omega_v=0} \end{cases},
\end{equation}
where $\T\N$ denotes the truncated normal distribution. Let $\meta_{\lambda}=(\eta_{\lambda, 1},\ldots,\eta_{{\lambda, V}})'$ and $\meta_{\omega}=(\eta_{\omega, 1},\ldots,\eta_{{\omega, V}})'$, then:
\begin{equation}
\pi(\mdelta_{\lambda}\mid\meta_{\lambda}, \kappa_{\lambda})=\N_q\left(\frac{1}{\kappa_{\lambda}}\left(\mQ_{s}+\mM{'}\mM\right)^{-1}\mM{'}\meta_{\lambda}, \quad \frac{1}{\kappa_{\lambda}}\left(\mQ_{s}+\mM{'}\mM\right)^{-1}\right),
\end{equation}
\begin{equation}
\pi(\mdelta_{\omega}\mid\meta_{\omega}, \kappa_{\omega})=\N_q\left(\frac{1}{\kappa_{\omega}}\left(\mQ_{s}+\mM{'}\mM\right)^{-1}\mM{'}\meta_{\omega}, \quad \frac{1}{\kappa_{\omega}}\left(\mQ_{s}+\mM{'}\mM\right)^{-1}\right).
\end{equation}
Moreover,
\begin{equation}
    \pi(\kappa_{\lambda}\mid\meta_{\lambda})
        =\G amma\left(a=\frac{V+1}{2},\quad b=\left[\frac{1}{2}\left(\sum_{v=1}^{V}\frac{\eta_{\lambda, v}^2}{(1+\mm_v{'}\mQ_{s}^{-1}\mm_v)}\right)+\frac{1}{2000}\right]^{-1}\right),
\end{equation}
\begin{equation}
    \pi(\kappa_{\omega}\mid\meta_{\omega})
        =\G amma\left(a=\frac{V+1}{2},\quad b=\left[\frac{1}{2}\left(\sum_{v=1}^{V}\frac{\eta_{\omega, v}^2}{(1+\mm_v{'}\mQ_{s}^{-1}\mm_v)}\right)+\frac{1}{2000}\right]^{-1}\right),
\end{equation}
where $b$ is the scale.

\end{document}